\documentclass[aip, jcp, reprint, twocolumn]{revtex4-1}
\usepackage{MnSymbol}
\usepackage{graphicx}
\usepackage{epstopdf}

\usepackage{graphics}
\usepackage{dcolumn}
\usepackage{bm}
 
\begin{document}
 
 
\title{Water above the spinodal} 
 
 
 
\author{Michal Du\v{s}ka}
\email[]{duskamich@gmail.com}
\affiliation{Institute of Thermomechanics of the CAS, v. v. i., Dolej\v{s}kova 1402/5, Prague 182 00, Czech Republic}
\affiliation{Institute for Physical Science and Technology, University of Maryland, College Park, Maryland 20742, USA}

 
\date{\today}
 
\begin{abstract}

The liquid spinodal, which is the bedrock of water thermodynamics, has long been discussed alongside the elusive liquid--liquid critical point hidden behind the limit of homogeneous nucleation. This has inspired numerous scenarios that attempt to explain water anomalies. Despite recent breakthrough experiments eliminating several of thous scenarios, we lacked a tool to localize the spinodal and the liquid--liquid critical point. We constructed a unique equation of state combining the famous Speedy's equation and the liquid--liquid critical point to remove that deficit and to review these explanations. For the first time, the proposed equation of state independently depicts the spinodal in the presence of the liquid--liquid critical point and demonstrates that the explanation for water anomalies based on the reentrance of the spinodal is not valid; this feature (reentering the spinodal) was solely predicted based on the curved density surface caused by the existence of the second critical point. However, the critical point alone is not sufficient to explain the shape of the density surface of water. In the new equation, hydrogen bond cooperativity is important to force the critical point to exist outside of zero temperature. Together with mounting evidence for the existence of a compressibility maximum behind the homogeneous nucleation limit at positive pressure, the findings practically exclude all explanations for water anomalies except for the existence of the liquid--liquid critical point at positive pressure. Finally, an extensive study of heat capacity demonstrated profound disagreement between the two major experimental heat capacity datasets and identified the more accurate dataset.

\end{abstract}

 
\maketitle 
 
 
\section{Introduction}
 
While considerable scientific effort has been devoted to discovering and proving the origin of water anomalies, conclusively determining the origin remains a difficult task. This paper presents an important step forward by demonstrating that all the proposed explanations for water anomalies can be captured by one phenomenon. As described below, our work builds upon decades of progress.
 
Water is a mysterious substance. Organisms that live in water owe their survival to the expansion of water upon freezing along with the fact that the maximum density of water occurs at 4$^\circ$C. The melting of water under pressure reduces the friction on the surface of ice, facilitating ice skating. This list can be expanded to include an astonishing 64 anomalies,\cite{WaterAnom} some of which define life as we know it.
 
The systematic effort to discover the origin of water anomalies is relatively new. Early experiments revealed rapid increases in the heat capacity\cite{Rasmussen1973a} and compressibility\cite{Kanno1979} of water upon supercooling to the point where no bulk experiments can be performed due to rapid freezing\cite{Kanno1975} (i.e., the homogeneous nucleation limit); however, a maximum or stability limit (where compressibility and heat capacity diverge) was not reached. After a decade of effort, Speedy reported an elegant way to analyze a different stability limit: the liquid spinodal.\cite{Speedy1982eos} Speedy discovered that the spinodal seems to turn back toward positive pressure, representing yet another possible anomaly. This finding indicated a connection between these two stability limits and gave rise to the first explanation for water anomalies: reentering the limit of stability. The Widom line\cite{GalloToKimScienc2017} was recently detected based on measurements of the compressibility and correlation length of evaporating microdroplets\cite{Kim2017Scienc} (the results are still controversial, primarily because the validity of the temperature measurement remains debated\cite{Caupin2018scienc, Kim2018sciencRes}) and the speed of sound in water in the double-metastable region\cite{Holten2017} (simultaneously supercooled and stretched). The detection of the Widom line is thought to indicate the existence of the liquid--liquid critical point (LLCP).\cite{Xu2005WdomLL} 
 
Well before the above experiments with microdroplets and stretched water, the phase transition between high-density and low-density water was discovered based on a molecular dynamics study, revealing the second critical point.\cite{Poole1992} The hypothetical LLCP then emerged as a second explanation for water anomalies. Additional studies were required to clarify whether low-density water is merely an ice-like fluctuation,\cite{Limmer2011,Limmer2013} or if an energy barrier exists between the new type of water and ice; a rigorous analysis\cite{Palmer2014aNature} demonstrated the existence of the energy barrier and phase transition. The LLCP explanation includes two special cases: (1) the critical point is hidden at zero temperature, representing a singularity-free scenario\cite{Sastry1996}; and (2) the critical point is hidden below the liquid spinodal, representing a critical point-free scenario.\cite{Angell2008} Thus, in addition to the existence of the critical point, the position of the critical point is also crucial.
 
Several cleverly designed physical experiments prevented freezing during supercooling to study the second critical point. Liu et al. studied confined water in nanoporous glass and found evidence of a fragile-to-strong dynamic transition in water.\cite{Liu2005} Meanwhile, by studying a crystallization-resistant aqueous solution, Woutersen et al. found evidence for the existence of a liquid--liquid transition in pure water.\cite{Woutersen2018} While both of these findings substantiate the existence of the critical point, these results cannot be directly linked to the properties of bulk water. Instead, the above studies clarify that the straightforward localization of the critical point and confirmation of the correct scenario is highly unlikely.\cite{Gallo2016} Herein lies the main difficulty in identifying the origin of water anomalies; thus, we must extrapolate toward the unknown in a compelling way.
 
Water is special because of its ability to form hydrogen bonds (HBs), which are relatively strong electromagnetic interactions between water molecules arising from polarity of the molecules. These HBs are strongly related to the orientation of the water molecules, which form a wide tetrahedral network (called tetrahedral water) that is responsible for decreasing the density of water.\cite{Taschin2013,Huang2009} HB strength and cooperativity have profound consequences.\cite{Stokely1301} Several studies have demonstrated the influence of HBs on the shape of the water phase diagram. For example, Sastry et al.\cite{Stastry1933LGHB} developed a lattice model that considers microscopic HB behavior to reproduce the anomalous behavior of water. Poole et al.\cite{Poole1994VdWHB} proposed an extension of the van der Waals equation of state (EOS) that accounts for the HB network in liquid water, and Chitnelawong et al.\cite{Chitnelawong2019VdWHB} reevaluated this extended EOS in consideration of the reentrant spinodal scenario. If tuned properly, these models can reproduce all the scenarios mentioned above. However, none of these models can describe the thermodynamic properties of water with sufficient accuracy; thus, these models cannot provide definitive proof for the origin of water anomalies.
 
The two-state method, which is fully thermodynamic (i.e., no knowledge of the microscopic behavior is required), can accurately reproduce water behavior.\cite{Holten2012,Russo2014NatCom} The International Association for the Properties of Water and Steam adopted a two-state EOS for supercooled ordinary water.\cite{Holten2014} The two-state model is based on the concept of a seemingly simple mixture or reaction between two states; however, it can also be seen as a first approximation of the bond lattice,\cite{Angell1971} which represents the complex behavior of HBs. Such a model can generate all the scenarios discussed above with the exception of reentering the limit of stability.\cite{Anisimov2018} In addition to implementing and testing the critical point scenario, Tanaka used the two-state method to rigorously examine the singularity-free scenario\cite{Tanaka2000}; however, subsequent density measurements reported by Mishima\cite{Mishima2010} indicated a liquid--liquid transition rather than a Widom line at elevated pressure, making a strong argument against the singularity-free scenario.

The main disadvantage of the two-state equation is that a polynomial function is used to approximate the high-density state of water (for simplicity, we refer to this state as water with broken HBs). Therefore, unlike the abovementioned models that consider hydrogen bonding, the core of the two-state equation does not have an explicit physical foundation. In the most advanced version of the two-state equation, the limit of stability is added to the polynomial function to represent the liquid spinodal.\cite{Biddle2017,Caupin2019} Herein, we propose a different approach based on the following considerations. If the line of compressibility maxima is real, the reentrance of the spinodal hypothesis cannot be valid. This does not mean that the reentrance feature is irrelevant quite the contrary. This hypothesis was proposed based on an elegant extrapolation of density from the limit of stability up to elevated pressure using a simple but accurate expansion. Thus, the curvature of the density surface that creates the appearance of reentrance is very real, and the theory capable of explaining this curvature will also be able to explain the nature of water anomalies.
 
Herein, we used the same equation proposed by Speedy for the extrapolation, constructed its integrated form that is able to describe all measurable thermodynamic properties, and implemented this form into the two-state equation. As will be discussed later, only a model consisting of components with clear physical meaning is capable of clarifying important issues. The new EOS can explain all proposed scenarios except for reentering the limit of stability, which was experimentally disproved by the existence of the compressibility maximum. The new EOS can also predict the spinodal for any of the remaining scenarios and explain the discovery of the seeming reentrance of the spinodal. Only experimental data can be used to determine which scenario is the most compelling to explain the origin of water anomalies.
 
\section{Equation of state}
 
To choose the proper form of the two-state model, the nature of the interaction between the two states must be examined. Molecular dynamics studies indicate that the mixing of states is symmetric.\cite{Singh2016,Russo2014NatCom} The HBs are apparently not strong enough to form a polymer-like supermolecule, in which the stoichiometry would create asymmetry. Thus a nonideal mixing contribution to the model will be fully responsible for the HB cooperativity.
 
The EOS in terms of reduced Gibbs free energy $\hat{G} = G/RT_{VLCP}$ (in reduced temperature $ \hat{T} = T/T_{\mathrm{VLCP}}$ and pressure $\hat{p} = p/p_{\mathrm{VLCP}}$, where VLCP indicates the conditions at the vapor--liquid critical point, and $R$ is the specific gas constant) is defined based on a nonideal mixture of tetrahedral water ($G^{\mathrm{B}}$) and water with broken HBs ($G^{\mathrm{A}}$):

\begin{eqnarray}
\hat{G}&&=\hat{G}^{\mathrm{A}}+
x \left( \hat{G}^{\mathrm{B}} - \hat{G}^{\mathrm{A}} \right) \nonumber\\
&& + \hat{T} \left[ x \mathrm{ln}x
+ \left( 1 - x \right) \mathrm{ln}\left( 1 - x \right)  \right] +
\omega x \left( 1 - x \right)
\label{eq:1},
\end{eqnarray}
where $x$ is the fraction of tetrahedral water, and $ \omega = \omega_{0}\left( 1 + \omega_{1}\hat{p} + \omega_{2}\hat{T} + \omega_{3}\hat{T}\hat{p} \right)  $  is the cooperativity. The variable $ \omega_{0} $ effectively serves as a LLCP parameter that ensures the critical point forms at the selected temperature.
 
The time required to reorganize a molecule of water (i.e., the structural relaxation time) is significantly lower than the time required for crystal nucleation and growth, even far from equilibrium.\cite{Limmer2011} Therefore, we can apply the equilibrium reaction conditions
 
\begin{eqnarray}
\left(\frac{\partial G}{\partial x}\right)_{T,P}=0
\label{eq:2}
\end{eqnarray}
to find a metastable value of $ x $ in all ranges of pressures and temperatures.
 
\subsection{Water as a simple fluid}
 
Speedy's EOS assumes that pressure expands above the limit of stability along an isotherm as a Taylor series of the difference between density and density at the spinodal.\cite{Speedy1982eos} If a second expansion term is applied, the volume of state A is given by
 
\begin{eqnarray}
\hat{V}_{\mathrm{A}} =
\left( \frac{\partial \hat{G}_{\mathrm{A}}}{\partial \hat{p}} \right)_{T} =
\frac{\hat{V}_{\mathrm{S}} \left( \hat{T}\right) B \left( \hat{T} \right)  }{\sqrt{1-\hat{p}/\hat{p}_{\mathrm{S}} \left( \hat{T}\right)} + B \left( \hat{T} \right)}
\label{eq:3},
\end{eqnarray}
where $ \hat{p}_{\mathrm{S}} $ is the pressure at the spinodal as a function of temperature, $ \hat{V}_{\mathrm{S}} $  is the volume at the spinodal,
 
\begin{eqnarray}
B = \sqrt{-\frac{\left(\frac{\partial ^2\hat{p}}{\partial \hat{\rho}^2}\right)_{T, sp.}}{2 \hat{p}_{\mathrm{S}} \hat{V}^2_{\mathrm{S}}}},
\end{eqnarray}
and $\left(\frac{\partial ^2\hat{p}}{\partial \hat{\rho}^2}\right)_{T, sp.}$ is the second derivative of pressure with respect to density at constant temperature at the spinodal. By definition, this second derivative is zero at the VLCP, which means that the equation has limited validity at elevated temperature because no higher term is included in Speedy's EOS. However, the EOS was demonstrated to well represent the density of water at temperatures up to 370~K and pressures from 0 to 100~MPa, and the addition of another expansion term did not considerably increase the quality of fit in the low-temperature region.\cite{Speedy1982eos}

State B is derived from the difference between the Gibbs free energies of states B and A:
 
\begin{eqnarray}
\hat{G}^{\mathrm{B}} - \hat{G}^{\mathrm{A}} &&=
a_{0} +
a_{1} \hat{p}\hat{T} +
a_{2} \hat{p} +
a_{3} \hat{T}
\nonumber\\
&& +
a_{4} \hat{T}^{2} +
a_{5} \hat{p}^{2} +
a_{6} \hat{p}^{3}
\label{eq:4},
\end{eqnarray}
 where the parameters $ a_{0} $, $ a_{2} $, $ a_{3} $... represent the first approximation of the difference between states B and A for $G$, $V$, $S$, and so on.
 
In this stage, by knowing the cooperativity and $ \hat{G}^{\mathrm{B}} - \hat{G}^{\mathrm{A}} $, it is possible to determine all three parameters in Speedy's EOS: the location of the spinodal in the phase diagram $ \hat{p}_{\mathrm{S}} $; the volume at the spinodal $ \hat{V}_{\mathrm{S}} $; and the second derivative $\left(\frac{\partial ^2\hat{p}}{\partial \hat{\rho}^2}\right)_{T, sp.}$. This represents the first approximation of the real spinodal based only on a model; since Caupin and Anisimov inserted the spinodal into the model, their results do not represent an independent estimate of the spinodal.\cite{Caupin2019}
 
To develop a fully thermodynamic description, we used polynomials to represent the temperature-dependent parameters of the spinodal:
 
\begin{eqnarray}
&&\hat{p}_{\mathrm{S}} =
1 +
d_{1} \left(\hat{T}-1\right) +
d_{2} \left(\hat{T}-1\right)^{2} +
d_{3} \left(\hat{T}-1\right)^{3}
\nonumber\\
&&\hat{V}_{\mathrm{S}} =
c_{0} +
c_{1} \left(\hat{T}\right) +
c_{2} \left(\hat{T}\right)^{2} +
c_{3} \left(\hat{T}\right)^{3} +
c_{4} \left(\hat{T}\right)^{4}
\\
&&\left(\frac{\partial ^2\hat{p}}{\partial \hat{\rho}^2}\right)_{T, sp.} =
b_{0} +
b_{1} \left(\hat{T}\right) +
b_{2} \left(\hat{T}\right)^{2} + b_{3} \left(\hat{T}\right)^{3} +
b_{4} \left(\hat{T}\right)^{4} \nonumber
\label{eq:5}.
\end{eqnarray}
We then constructed the integrated form of Speedy's EOS by determining the entropy of state A (which was more convenient then deriving the Gibbs free energy itself) from the Maxwell relation $ \left( \frac{\partial S}{\partial V}  \right)_{T} = \left( \frac{\partial p}{\partial T} \right)_{V}  $, where the right side comes from Eq.~(\ref{eq:3}):
 
\begin{eqnarray}
\hat{S}_{\mathrm{A}} && \left( \hat{V}_\mathrm{A} \left(\hat{p}, \hat{T} \right), \hat{T} \right) =
- \left( \frac{\partial \hat{G}_{\mathrm{A}}}{\partial \hat{T}} \right)_{p}  \nonumber\\
&& = \frac{d \hat{p}_{\mathrm{S}}}{d \hat{T}} \left( \hat{V}_{\mathrm{A}} - \hat{V}_{\mathrm{S}} \right)
+ A \left(\hat{T} \right)\left[\frac{\hat{V}_{\mathrm{S}}^{2} }{\hat{V}_{\mathrm{A}}}  + 2 \hat{V}_{\mathrm{S}}\mathrm{ln}\frac{\hat{V}_{\mathrm{A}}}{\hat{V}_{\mathrm{S}}} - \hat{V}_{\mathrm{A}}
\right]
\nonumber\\
&& + C \left(\hat{T} \right)\left[\frac{\hat{V}_{\mathrm{S}} }{\hat{V}_{\mathrm{A}}}  + \mathrm{ln}\frac{\hat{V}_{\mathrm{A}}}{\hat{V}_{\mathrm{S}}} - 1
\right]
+ \hat{S}_{\mathrm{S}} \left( \hat{T} \right)
\label{eq:6},
\end{eqnarray}
where $ A =\frac{d \hat{p}_{\mathrm{S}}}{d \hat{T}} B^{2} + 2p_{S}B\frac{d B}{d \hat{T}} $, $ C =2p_{S}B^{2}\frac{d \hat{V_{S}}}{d \hat{T}}$, and $\hat{S}_{\mathrm{S}}$ is the entropy at the spinodal, which is yet to be determined.
 
\subsection{Optimization of EOS parameters}

As a first step in EOS development, the parameters of the two-state model were optimized to fit the density at positive pressure and to predict the parameters of Speedy's EOS at selected temperatures. A spline representation of the spinodal was introduced to evaluate the accuracy of the speed of sound prediction. This relatively fast optimization served as a global optimization tool to find the proper two-state model. As a second step, the parameters of the two-state model and the analytical representation of state A were simultaneously optimized to fit the density via genetic optimization. Subsequently, the derivative of the entropy of state A at the spinodal $ \hat{S}_{\mathrm{S}}^{\prime} = \frac{d \hat{S}_{\mathrm{S}}}{d \hat{T}} $ was introduced based on the heat capacity at atmospheric pressure. By selecting the proper analytical representation, we included the heat capacity and speed of sound into the optimization to further improve the model. There is an important issue to resolve: the major datasets of heat capacity in the supercooled region (Angell et al.,\cite{Angell1982Cpbest} Tombari et al.,\cite{TOMBARI1999} and Voronov et al.\cite{Voronov2018}) disagree significantly about the speed at which heat capacity increases with decreasing temperature. 

For $ \hat{S}_{\mathrm{S}}^{\prime} $, we choose the following simple cubic polynomial to impose a restriction on the continuous character of $ \hat{S}_{\mathrm{S}}^{\prime} $:

\begin{eqnarray}
\hat{T}\hat{S}_{\mathrm{S}}^{\prime} =
s_{0} +
s_{1} \left(\hat{T}\right) +
s_{2} \left(\hat{T}\right)^{2} +
s_{3} \left(\hat{T}\right)^{3}
\label{eq:7}.
\end{eqnarray}
 
\begin{figure*}
 \includegraphics[scale=1]{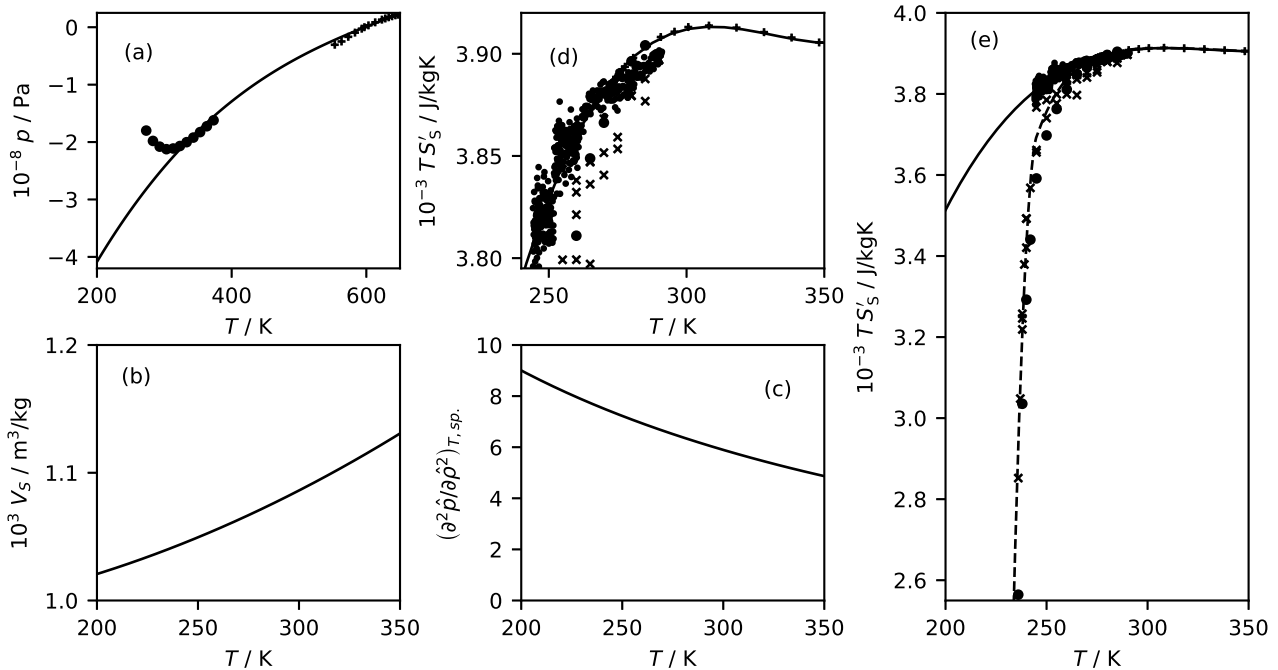}    
 \caption{\label{fig:wide1} The five temperature-dependent functions of the EOS that best represent the heat capacity of Tombari et al.\cite{TOMBARI1999}/Voronov et al.\cite{Voronov2018} (solid line) and Angell et al.\cite{Angell1982Cpbest} (dashed line). (a) The spinodal pressure of state A, where $\bullet$ symbols represent the pressure at the spinodal determined by Speedy,\cite{Speedy1982eos} and $+$ symbols represent those of Chukanov and Skripov.\cite{ChukanovSkripov1971} (b) Volume at the spinodal. (c) Second derivative of pressure with respect to density at constant temperature at the spinodal. (d, e) Derivative of the entropy of state A at the spinodal. The heat capacity data at atmospheric pressure were recalculated to their values at the spinodal to be fitted with the derivative of entropy. The $ \cdot$ symbols represent the data of Voronov et al.,\cite{Voronov2018} the $\times $ symbols are the data of Angell et al.,\cite{Angell1982Cpbest} the $ \bullet$ symbols are the data of Archer and Carter,\cite{Archer2000} and the $ +$ symbols are the data of Osborne et al.\cite{Osborne1939Cp} }
\end{figure*}

It was possible to optimize the model based only on the data of Tombari et al.[Editor7]\cite{TOMBARI1999} and Voronov et al.\cite{Voronov2018} (Table~\ref{tab:table1}; EOS-VaT model) but not based on the data of Angell et al.\cite{Angell1982Cpbest} or Archer and Carter\cite{Archer2000} (whose heat capacity is growing even more slowly then Angell et al.). To fit these data, $ T \, S_{\mathrm{S}}^{\prime} $ must decrease sharply (see Fig.~\ref{fig:wide1}). Such a rapid change in $ T \, S_{\mathrm{S}}^{\prime} $ independent of pressure at the same temperature for both high- and low-density water is highly unlikely and would require considerations beyond hydrogen bonding. Based on experiments in water-rich hydrazinium trifluoroacetate solution, Woutersen et al.\cite{Woutersen2018} found that the sharp change in heat capacity can be attributed solely to the change in HB structure, indicating proximity to the liquid--liquid transition. To demonstrate the thermodynamic consequences of this behavior, we accepted the existence of such a change and used spline interpolation as an analytic representation of $\hat{T}\hat{S}_{\mathrm{S}}^{\prime} \left(\hat{T}\right)$ for the model of Angell et al.\cite{Angell1982Cpbest} (EOS-A model). If asked, the author can provide the parameters of this model.
 
\begin{table*}
\caption{\label{tab:table1}Parameters of the EOS-VaT model based on the data of Voronov et al.\cite{Voronov2018} and Tombari et al.\cite{TOMBARI1999}}
\begin{ruledtabular}
\begin{tabular}{ccccccc}
& $a$ & $b$ & $c$ & $d$ & $\omega$ & $s$\\
\hline
0 & -4.3743227e-01 & 1.0783316e+02 & 7.7905108e-03 & - & 4.1420925e-01 & 1.1210116e+02\\
1 & -1.3836753e-02 & -2.3582144e+02 & -1.3084560e-02 & 1.2756957e+01 & 3.6615174e-02 & -1.7214704e+02\\
2 & 1.8525106e-02 & 2.1364435e+02 & 5.7261662e-02 & 2.8548221e+01 & 1.6181775e+00 & 8.7732559e+01\\
3 & 4.3306058e-01 & -1.0405443e+02 & -8.9096098e-03 & -2.6960321e+00 & 7.1477190e-03 & -6.3674996e+00\\
4 & 2.1944047e+00 & 2.6732998e+01 & 7.3009898e-02 & - & - & -\\
5 & -1.6301740e-05 & - & - & - & - & -\\
6 & 7.6204693e-06 & - & - & - & - & -\\
 
\end{tabular}
\end{ruledtabular}
\end{table*}
However, the independent optimization of the EOS-VaT and EOS-A models did not produce a significantly different state A or two-state model. Therefore, the five temperature-dependent functions presented in Fig.~\ref{fig:wide1} correspond to the EOS-VaT model (solid line); the dashed line corresponds to the  $\hat{T}\hat{S}_{\mathrm{S}}^{\prime} \left(\hat{T}\right)$ of the EOS-A model. This means that both models are identical with the exceptions of heat capacity and properties derived from heat capacity as speed of sound, for which the differences between models are insignificant. The pressure at the spinodal $ \hat{p}_{\mathrm{S}} \left( \hat{T}\right) $  connects smoothly to Speedy's limit of stability at higher temperatures [see Fig.~\ref{fig:wide1}(a)], suggesting the vanishing of tetrahedral water. The volume at the spinodal $ \hat{V}_{\mathrm{S}} \left( \hat{T}\right) $ [Fig.~\ref{fig:wide1}(b)] is a monotonic function whose slope gradually decreases with decreasing temperature. The second derivative of pressure with respect to density at constant temperature at the spinodal [Fig.~\ref{fig:wide1}(c)] behaves according to theory; that is, it decreases toward zero as the VLCP is approached. The derivative of the entropy of state A at the spinodal $ \hat{T} \hat{S}_{\mathrm{S}}^{\prime} $  indicates the sharp distinction between the two heat capacity datasets.
 
An EOS based on only one expansion term with respect to the pressure in state A cannot be accurate over large pressure ranges. Of the 28 fitting parameters of the entire EOS, 17 are the parameters of Speedy's EOS and the temperature-dependent entropy at the spinodal. However, these 17 parameters only represent four physical quantities for which accurate temperature representation is crucial. Thus, the only parameters of the EOS are the position of the spinodal and its thermodynamic properties, the difference between the thermodynamic properties of states A and B, and the cooperativity. Determining these properties without interference from the polynomial terms in state A, which have no clear physical meaning, is critical for identifying the origin of water anomalies.
 
\section{Experimental data}
\begin{figure}
 \renewcommand{\arraystretch}{0}
 \includegraphics[scale=1]{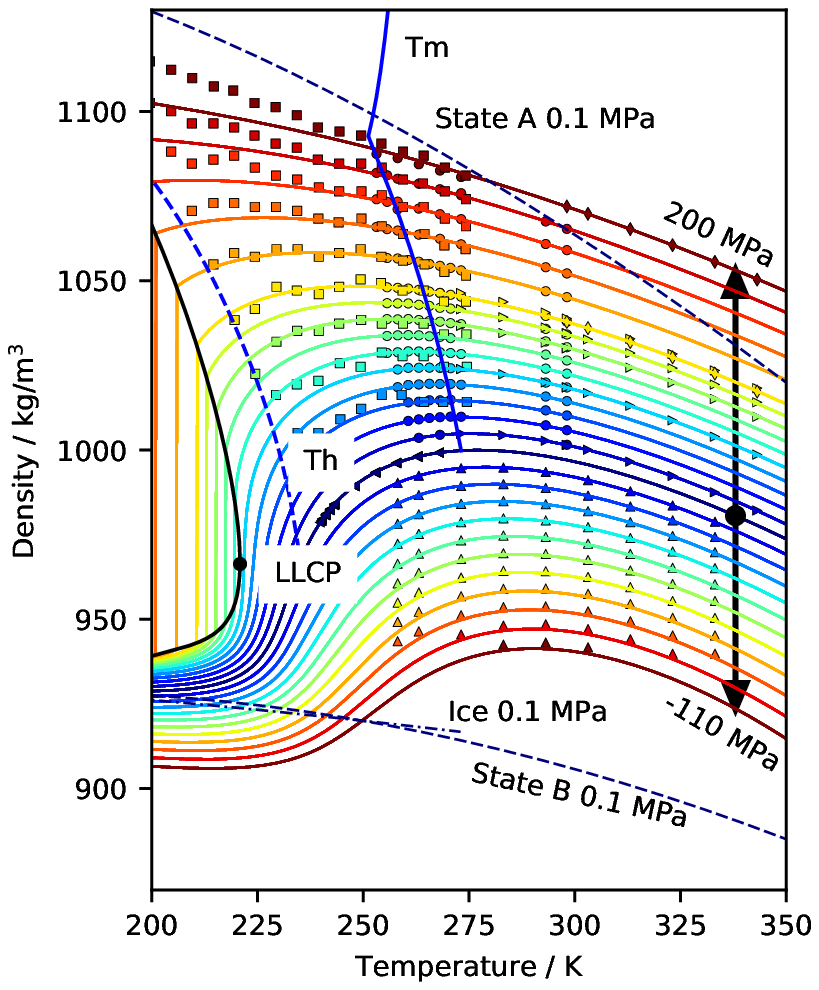}
 \caption{\label{fig:wide2} Density of the EOS-VaT model at pressures between $-110$~MPa and 100~MPa in 10-MPa intervals and then at 20-MPa intervals up to 200~MPa in comparison to the experimental data of Mishima\cite{Mishima2010} ($\blacksquare$), Hare and Sorensen\cite{Hare1987} ($\blacktriangleleft$), Sotani et al.\cite{Sotani2000} ($\bullet$), Kell and Whalley\cite{Kell1975} ($\blacktriangleright$), Grindley and Lind\cite{Grindley1971} ($\blacklozenge$), and Pallares et al.\cite{Pallares2016} ($\blacktriangle$). The included experimental pressures deviate by less than 3\% from the model results. The solid blue line marked with Tm is the density of water at the melting point.\cite{Feistel_Wagner2006Ih} The dashed blue line marked with Th is the density at the homogeneous nucleation limit.\cite{Kanno1975} The densities of states A and B at atmospheric pressure are shown as dashed lines. The density of Ih ice at atmospheric pressure\cite{Feistel_Wagner2006Ih} is shown as a dot--dash line, and the LLTL ending at the critical point is shown as a solid black line with a circle labeled LLCP.}
\end{figure}

It is important to point out that the proposed EOS aims to be valid at pressures up to 100~MPa; however, we discuss data for pressures up to 200~MPa to ensure proper extrapolation ability. The selected thermodynamic data represent the state of the art; the particular datasets are indicated in the figure captions. We have already shown that our model predicts normal-like spinodal behavior without the spinodal reentrance feature. We now show how well the model can explain the curvature of the density surface (Fig.~\ref{fig:wide2}). At positive pressure, we relied on density measurements of bulk and emulsified water. For negative pressure, the density was derived from the speed of sound measurements reported by Pallares et al.\cite{Pallares2016} Thus, the measured density surface of ordinary water is a complete picture extending far below the temperature of freezing to 200~K and also down to pressures as low as $-110$~MPa. However, other conditions must be met to obtain a meaningful model. Throughout the optimization, the proximity of the density of state B to the density of tetrahedral ice at atmospheric pressure was required to match the observed proximity in Palmer et al.'s molecular model of water.\cite{Palmer2014aNature} To obtain a continuous pathway to amorphous ice, it is important for the liquid--liquid transition to have the proper shape; this proved to be a nontrivial problem. In two versions of the previously published EOS,\cite{Caupin2019} either density close to the liquid--liquid transition line (LLTL) at low pressure or the extrapolation toward high-density amorphous ice was poorly represented. Our EOS successfully reproduces both; however, because a cubic term was added to $ \hat{G}^{\mathrm{B}} - \hat{G}^{\mathrm{A}} $ to increase the flexibility of the LLTL and the Widom line, state B exhibits instability below 200~K (not visible in Fig.~\ref{fig:wide2}), resulting in improper behavior in the small region below the LLTL. 

On the other hand, the density at low temperatures and pressure above 100~MPa (just above the LLTL) could be improved by increasing the density at the liquid--liquid transition (the approach used by Caupin and Anisimov\cite{Caupin2019}). However, to accurately represent high-density amorphous ice, this approach requires a sharp reduction in the rate at which the density increases at the LLTL. Otherwise, the LLTL is at risk of extending far beyond the homogeneous nucleation limit,\cite{Caupin2019} which would contradict experimental data.\cite{Mishima2010} This issue could be addressed by including additional terms into $ \hat{G}^{\mathrm{B}} - \hat{G}^{\mathrm{A}} $, although that would further complicate the model. Thus, we consider our EOS to be a reasonable compromise, especially considering that the major focus is on the spinodal.
 
\begin{figure}
 \renewcommand{\arraystretch}{0}
 \includegraphics[scale=1]{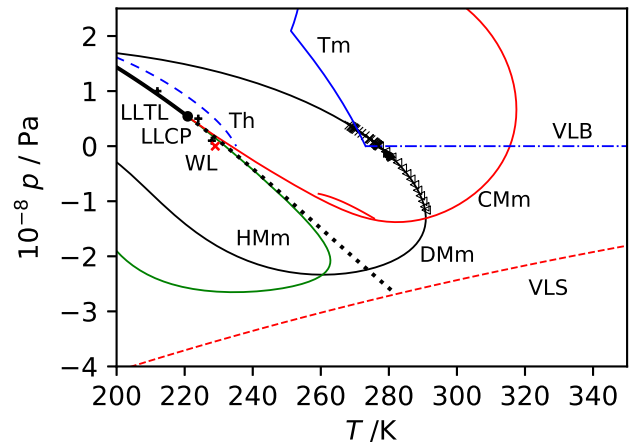}
 \caption{\label{fig:wide7} Phase diagram of the EOS-VaT model showing the density extrema locus (DMm; black curve), compressibility extrema locus (CMm; solid red curve), isobaric heat capacity extrema locus (HMm; green curve), Widom line (WL; dotted black line), vapor--liquid bimodal (VLB; blue dot--dash line), vapor--liquid spinodal (VLS; red dashed line), and melting line (Tm; blue dashed line). The DMm is compared to the experimental data of Caldewll\cite{Caldwell1978} ($ \times $), Henderson and Speedy\cite{Henderson1987} ($\blacktriangleright$), and Pallares et al.\cite{Pallares2016} ($\lhd$). CMm is compared with the data of Kim et al.\cite{Kim2017Scienc} (red $ \times$) and Holten et al.\cite{Holten2017} (short red solid line). A projection of the mechanical stability limit of Kanno et al.\cite{Kanno1975} is also presented ($+$).}
\end{figure}

\begin{figure}
 \renewcommand{\arraystretch}{0}
 
 \includegraphics[scale=1]{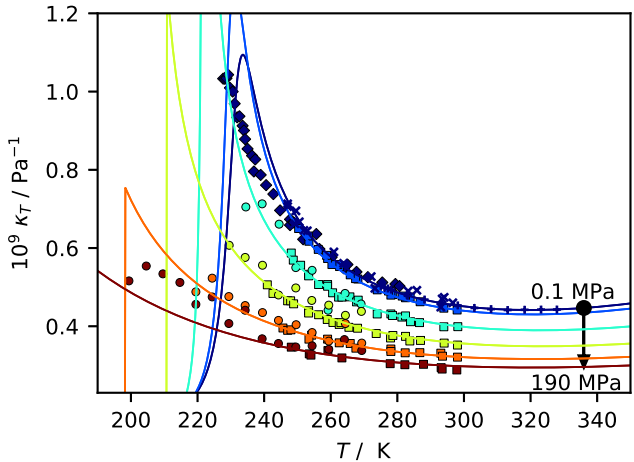}
 \caption{\label{fig:wide6} Compressibility of the EOS-VaT model  for pressures of 0.1~MPa (together with zero pressure), 10~MPa, 50~MPa, 100~MPa, 150~MPa, and 190~MPa. The experimental data of Kanno and Angell\citep{Kanno1979} ($ \blacksquare$), Millero et al.\citep{Millero1969} ($ +$), Mishima\citep{Mishima2010} ($ \bullet$), and Speedy and Angell\citep{Speed1976} ($ \times$) are shown for comparison. The model pressure deviates from the experimental data by less than 3\%.}
\end{figure}

The phase diagram shown in Fig.~\ref{fig:wide7} depicts the complex shape of the LLTL and the Widom line even better then density. In the phase diagram, the LLTL is hidden behind the homogeneous nucleation limit in the region of interest, in agreement with the experimental data.\cite{Mishima2010} However, the LLTL does not respond quickly enough to decreasing temperature to reflect the amorphous ice transition at 200~MPa. On the other hand, instead of decreasing monotonically with increasing temperature, the Widom line creates an S shape, which is crucial to properly interpret the following data. The locus of density and compressibility maxima measured by the group of Caupin\cite{Pallares2016, Holten2017} are correctly represented even though their data were not used for optimization. However, a slight disagreement of as much as 4.5~K still exists with respect to the position of the compressibility maximum determined by Kim et al.\cite{Kim2017Scienc} However, it is difficult to reconcile the trend in Kim et al.'s data with the data of Kanno and Angell,\citep{Kanno1979} as pointed out by Caupin and Anisimov\cite{Caupin2019} and shown in Fig.~\ref{fig:wide6}; thus, the precise position of this compressibility maximum may not be definitively established. The flexibility of the LLTL and the Widom line is crucial, as demonstrated by the case without such flexibility; without the cubic term in $ \hat{G}^{\mathrm{B}} - \hat{G}^{\mathrm{A}} $, it was impossible to properly reconcile these two critical datasets.\cite{Caupin2019} Thus, adding the cubic term to $ \hat{G}^{\mathrm{B}} - \hat{G}^{\mathrm{A}} $ is clearly reasonable.
 
After obtaining the basic shape of the LLTL and Widom line, we focused on the cooperativity. Once the criticality is introduced, higher terms are responsible for changing the fraction of conversion $x$, which significantly shapes the density surface. To capture the delicate shape of the density surface of water, the cooperativity $ \omega$ must be dependent on both pressure\cite{Holten2012} and temperature, thus adding two extra parameters to the model. This result is not surprising given the evidence supporting the importance of HB cooperativity, as mentioned above. The importance of cooperativity makes a strong argument against the singularity-free scenario; if the density surface cannot be explained without cooperativity, the LLCP must also be present.\cite{Tanaka2000}
 
Although an important parameter of the model, the location of the LLCP is approximate because the two-state model represents a mean-field approximation. We localized the LLCP at 220.9~K and 54.2~MPa; thus, the critical pressure is slightly less than the minimum of 60~MPa expected by Ni et al. based on a recent study of water diffusion.\cite{Ni2018} This is likely due to the insufficient flexible of LLTL and Widom line, as discussed above. The position of the LLCP at lower pressure then expected \cite{Ni2018} leads to higher compresibility maximum (Fig.~\ref{fig:wide6}) then measured at zero pressure \cite{Kim2017Scienc} confirming consistency of those important data sets and our model.
 
\begin{figure}
 \renewcommand{\arraystretch}{0}
 \includegraphics[scale=1]{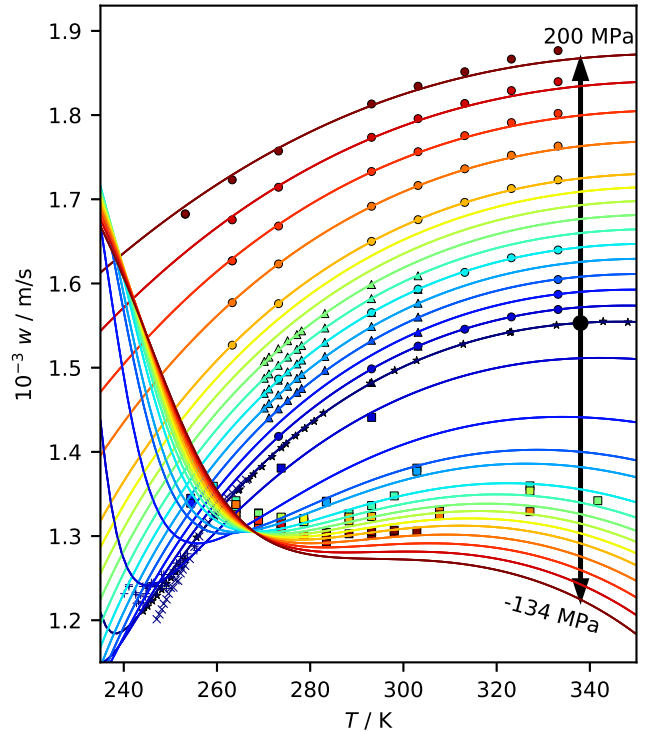}
 \caption{\label{fig:wide5} Speed of sound in the EOS-VaT model for pressures between 0.1 and  100~MPa in 10-MPa intervals and then at 25-MPa intervals up to 200~MPa as well as for negative pressures of $-21.5$, $-55$, $-73$, $-80.5$, $-91$, $-97$, $-102$, $-106$, $-110$, $-114$, $-119$, $-124$, $-129$, and $-134$~MPa. The model results are compared to the experimental data of Lin and Trusler\cite{LinTrusler2012} ($ \bullet$), Aleksandrov and Larkin\cite{AleksandrovLarkin1976} ($ \blacktriangle$), Taschin et al.\cite{Taschin2011} ($ \star$), Bacri and Rajaonarison\cite{BacriRajaonarison1979} ($ \times$), Trinh and Apfel\cite{TrinhApfel1980} ($ +$), and Holten et al.\cite{Holten2017} ($ \blacksquare$). The deviation between the model and experimental pressures is less than 3\% at positive pressure and 1~MPa at negative pressure.}
\end{figure}

To further scrutinize the model, we analyzed the speed of sound. The model precisely reproduces the experimental data at positive pressures up to 100~MPa (see Fig.~\ref{fig:wide5}) but gradually begins to disagree with the data at negative pressure. Without changing the model itself, we evaluated the possibility of improving the accuracy at negative pressure by including negative pressure data into the optimization; however, increasing the accuracy at negative pressure was only achieved at the expense of the accuracy at positive pressure. This suggests a difference in the rate of change in density between positive and negative pressures. We decided to retain the correct interpretation of positive pressure and use the negative pressure data as an indication of the model's extrapolation ability. Thus, the final EOS is optimized only for data at positive pressure.

\begin{figure*}
 \renewcommand{\arraystretch}{0}
 \begin{tabular}{cc}
   \includegraphics[scale=1]{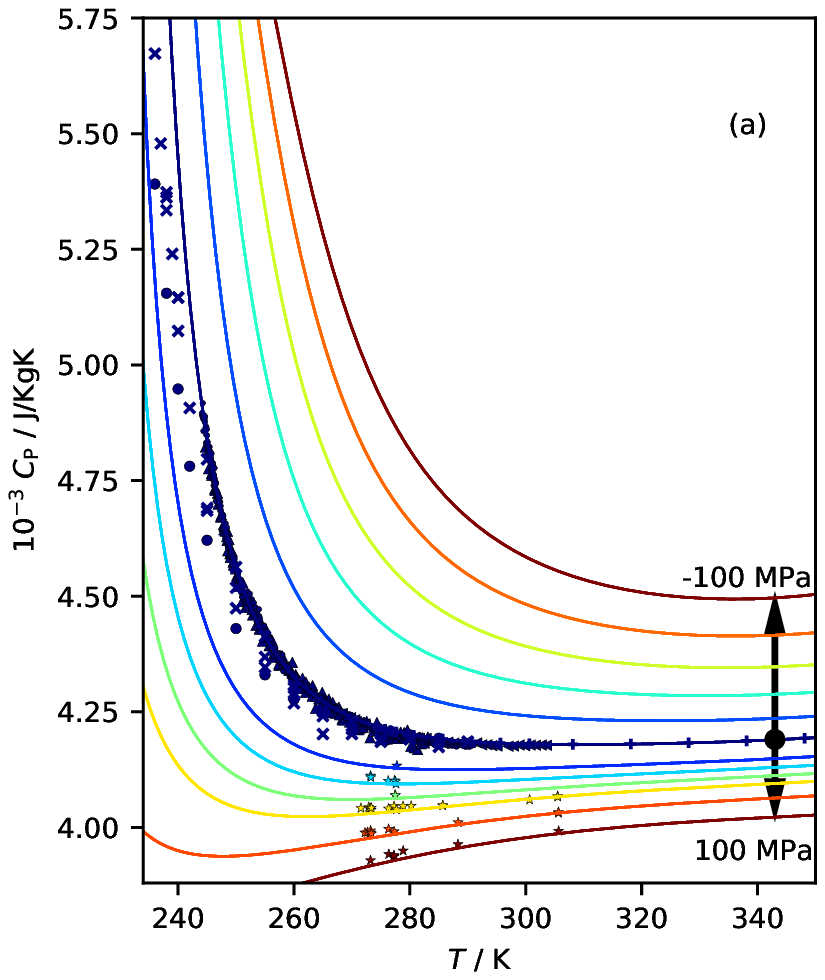} &
   \includegraphics[scale=1]{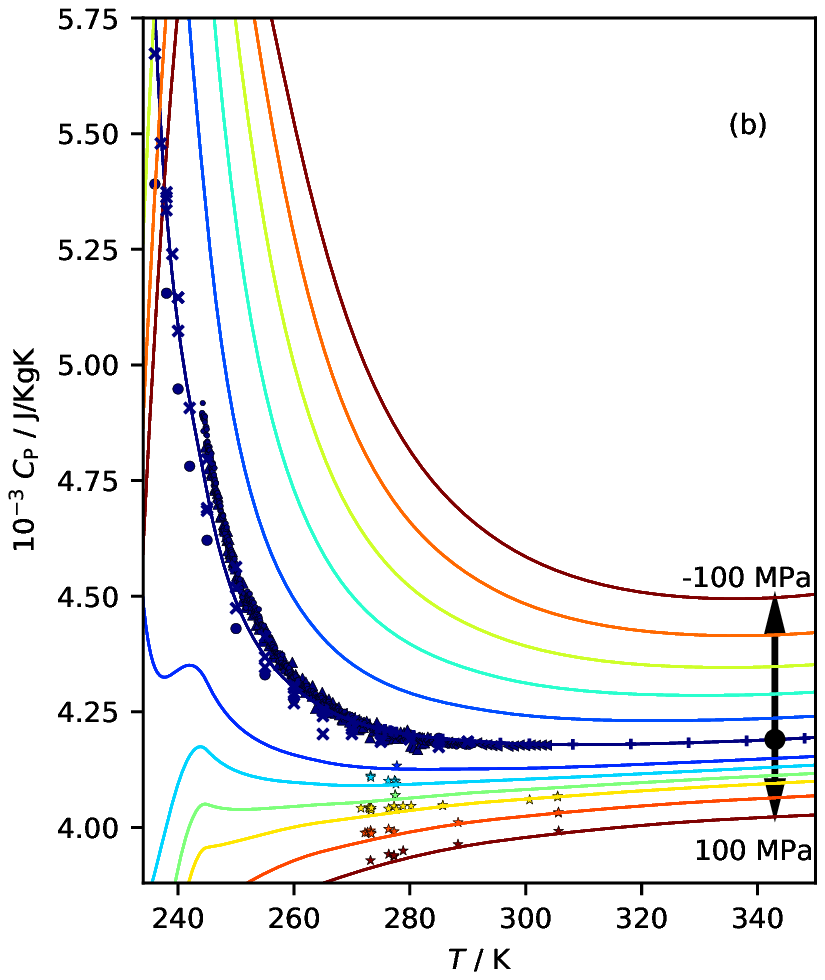}
 \end{tabular}
 \caption{\label{fig:wide3} Isobaric heat capacities of the EOS-VaT model (a) and the EOS-A model (b) for pressures between 0.1 and $-100$~MPa in intervals of 20 MPa and at 20, 30, 40, 50, 70, and 100~MPa. The heat capacities are compared to the experimental data of Angell et al.\cite{Angell1982Cpbest} ($ \times$), Voronov et al.\cite{Voronov2018} ($ \cdot$), Tombari et al.\cite{TOMBARI1999} ($ \blacktriangleleft$), Archer and Carter\cite{Archer2000} ($ \bullet$), Osborne et al.\cite{Osborne1939Cp} ($ +$), Sirota et al.\cite{Sirota1970} ($ \star$), and Anisimov et al.\cite{Anisimov1972} ($ \blacktriangleleft$). The deviation between the model and experimental pressures is less than 3\%.}
\end{figure*}

\begin{figure*}
 \renewcommand{\arraystretch}{0}
 \begin{tabular}{cc}
   \includegraphics[scale=1]{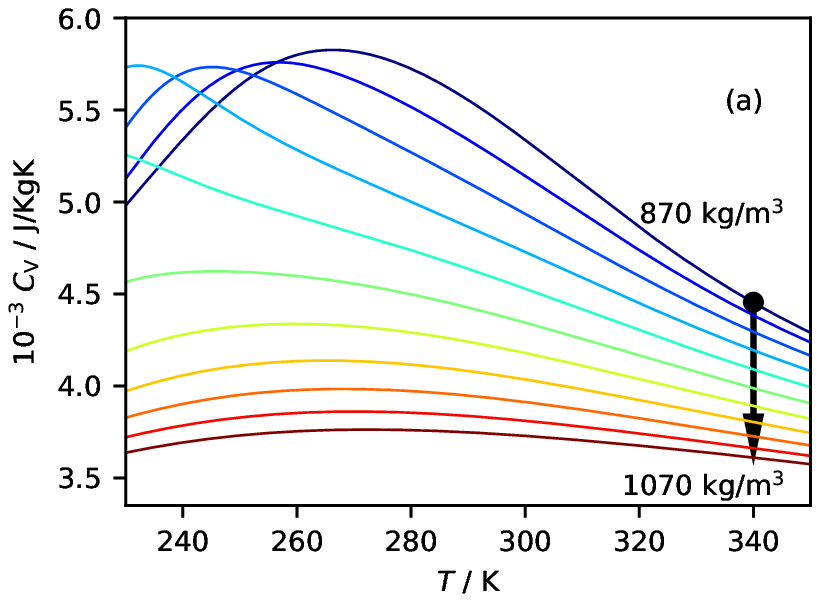} &
   \includegraphics[scale=1]{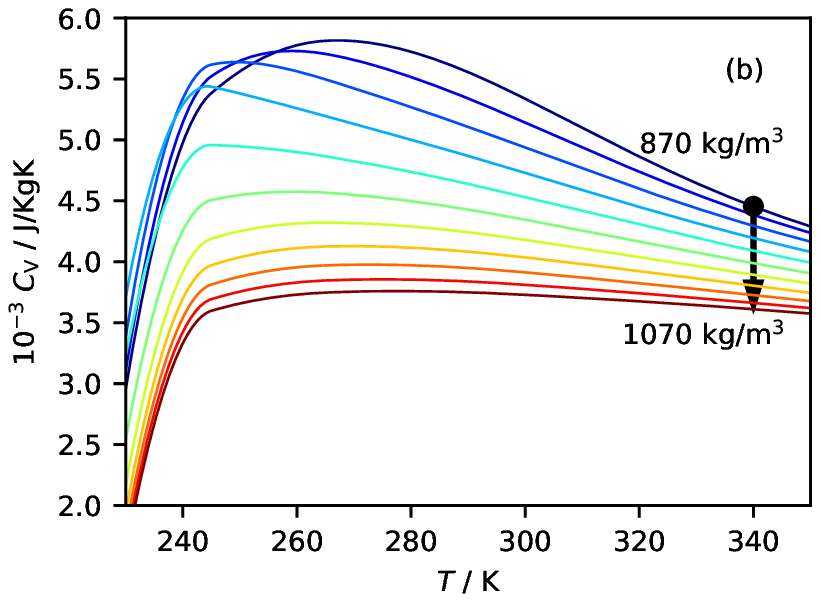}
 \end{tabular}
 \caption{\label{fig:wide4} Isochoric heat capacities of the EOS-VaT model (a) and the EOS-A model (b) for densities between 870 and 1070~kg/m$ ^{3}$ in intervals of 20~kg/m$ ^{3}$. }
\end{figure*}

As required by the definition of the model, the model accurately represents the isobaric heat capacity at atmospheric pressure. Figure~\ref{fig:wide3}(b) shows one consequence of the sharp decrease in $ \hat{T} \hat{S}_{\mathrm{S}}^{\prime} $ necessary to fit the data of Angell et al.\cite{Angell1982Cpbest} If not compensated by an increase in heat capacity due to the proximity of the Widom line, this decrease in $ \hat{T} \hat{S}_{\mathrm{S}}^{\prime} $ leads to a decrease in the heat capacity itself at a pressure slightly above atmospheric pressure. This phenomenon is even more clearly exhibited by the isochoric heat capacity [Fig.~\ref{fig:wide4}(b)]. Such a decrease in heat capacity is not found in the molecular dynamics model of water reported by Singh et al.\cite{Singh2016} On the other hand, the EOS-VaT model [see Fig.~\ref{fig:wide4}(a)] is similar to Singh et al.'s molecular model of water, further confirming that heat capacity data of Tombari et al.\cite{TOMBARI1999} and Voronov et al.\cite{Voronov2018} better describe the trend in heat capacity. Figure~\ref{fig:wide3}(a) also demonstrates good agreement between the EOS and the measured data at elevated pressure.
 
\section{Conclusion}
The curious thermodynamic properties of water have inspired the investigation of vast regions below the freezing point and even regions of negative pressure. These investigations have led to several explanations for the anomalous behavior of water. Evidence suggests that the compressibility maximum is located below the homogeneous nucleation limit at positive pressures.\cite{Kim2017Scienc} This means that only two of the four previously proposed scenarios can be true: the second critical point scenario and the singularity-free scenario. To determine which one of these scenarios is the most probable and explain why spinodal reentrance was postulated in the first place, we formulated a new EOS by combining the two-state model with Speedy's EOS in the same form that suggested the reentrance of the spinodal. For the first time, we demonstrated that the spinodal reentrance scenario arose from the existence of the LLCP at positive pressure, resulting in the curvature of the density surface. If the critical point is considered, our EOS predicts the spinodal without any anomalies. Despite its deliberate simplicity, the proposed EOS can describe all available thermodynamic data with reasonable accuracy and approximately locate both the second critical point and the spinodal, which are both of immense interest. We also demonstrated that the singularity-free scenario, which requires HB cooperativity to be negligible, allowing the critical point to hide in zero temperature, clearly contradicts the available data. Thus, despite a lack of direct evidence for the existence of the LLCP, no other plausible explanation remains. This elusive second critical point of water is the only explanation we can offer to explain the nature of water, a substance that shapes all life in a profound way.
 
The main advantage of the proposed EOS is its ability to predict the spinodal in the presence of the second critical point. This feature is a critical advantage of our EOS compared to the previously published model.\cite{Caupin2019} Another advantage of our EOS is its simplicity and clarity of all components, which allow the determination of the important parts of the model. The simplicity arises from the lack of additional polynomial terms that create artificial distortion between the two-state model and extrapolation toward the spinodal. The cooperativity seems to be more important than anticipated based on previous two-state models. The new EOS denotes water as a nonregular solution in which phase separation is driven by both excess energy and entropy. The most compelling way to reconcile the latest data on the compressibility maximum\cite{Holten2017,Kim2017Scienc} is to introduce greater flexibility into the curve forming the LLTL and the Widom line via the addition of an inflection point before crossing the spinodal. We also showed that the difference between the heat capacities measured by Voronov et al.\cite{Voronov2018} and Tombari et al.\cite{TOMBARI1999} and those reported by Angell et al.\cite{Angell1982Cpbest} is so profound that there is no rational basis for a strong change in heat capacity with temperature. According to our EOS, the measurements of Voronov et al.\cite{Voronov2018} and Tombari et al.\cite{TOMBARI1999} more accurately represent heat capacity in the metastable region. Our EOS also ensures reasonable extrapolation towards amorphous ice, opening a route towards complete EOS of polymorphous mater. That is the goal of our future effort.

\begin{acknowledgments}
We acknowledge Mikhail A. Anisimov for suggesting this research project and for providing new experimental data on the heat capacity of supercooled water, F. Caupin for providing new data on the speed of sound at negative pressure, and J.V. Singers, A.H. Harvey, and J. Hruby for fruitful discussions. This research originated from the project Towards an IAPWS Guideline for the Thermodynamic Properties of Supercooled Heavy Water (preliminary results have been made public\cite{Duska2017eosCom}) under support from the Young Scientist IAPWS Fellowship. Continuation of the research was supported by the Institute of Thermomechanics, Czech Academy of Sciences (institutional support RVO:61388998).
\end{acknowledgments}


\section{AIP Publishing Data Sharing Policy}
The data that support the findings of this study are available from the corresponding author upon reasonable request.

%
 
 
\bibliography{Com2017}
 
\end{document}